\documentclass[aps,prb,reprint,amsmath,showpacs,showkeys]{revtex4-1}

\usepackage{graphicx}
\usepackage{dcolumn}
\usepackage{bm}
\usepackage{float}

\begin{document}

\title{Effect of pressure cycling on Iron: Signatures of an electronic instability and unconventional superconductivity}

\author{C. S. Yadav$^{1}$, G. Seyfarth$^{1,2}$, P. Pedrazzini$^{3}$, H. Wilhelm$^{4}$, R. \v{C}ern\'y$^{5}$, and D. Jaccard$^{1}$}

\affiliation{$^{1}$DPMC - University of Geneva, 24 Quai
Ernest-Ansermet, 1211 Geneva 4, Switzerland}
\affiliation{$^{2}$Laboratoire National des Champs Magn\'{e}tiques
Intenses, CNRS and UJF, 38042 Grenoble, France}
\affiliation{$^{3}$Lab. Bajas Temperaturas and Instituto Balseiro,
CNEA, 8400 S.C. de Bariloche, Argentina}
\affiliation{$^{4}$Diamond Light source Ltd, Chilton, Didcot,
Oxfordshire, OX11 0DE, United Kingdom}
\affiliation{$^{5}$Laboratory of Crystallography, DPMC -
University of Geneva, 24 Quai Ernest-Ansermet, 1211 Geneva 4,
Switzerland}

\begin{abstract}



High pressure electrical resistivity and x-ray diffraction
experiments have been performed on Fe single crystals. The
crystallographic investigation provides direct evidence that in
the martensitic $bcc \rightarrow hcp$ transition at 14 GPa the
$\lbrace 110\rbrace_{bcc}$ become the $\lbrace 002\rbrace_{hcp}$
directions. During a pressure cycle, resistivity shows a broad
hysteresis of 6.5 GPa, whereas superconductivity, observed between
13 and 31 GPa, remains unaffected. Upon increasing pressure an
electronic instability, probably a quantum critical point, is
observed at around 19 GPa and, close to this pressure, the
superconducting $T_{c}$ and the isothermal resistivity
($0<T<300\,$K) attain maximum values. In the superconducting
pressure domain, the exponent $n = 5/3$ of the temperature power
law of resistivity and its prefactor, which mimics $T_{c}$,
indicate that ferromagnetic fluctuations may provide the glue for
the Cooper pairs, yielding unconventional superconductivity.

\end{abstract}


\maketitle

\section{Introduction}

The advent of superconductivity in the hexagonal phase of iron
between 13 and 31 GPa, described by Shimizu \textit{et al.} in
2001, was a surprise for the scientific community.\cite{Shimizu}
Despite the interest in this discovery, little experimental work
has been done so far.\cite{Jaccard, Alex, Pablo, Koshik} Given the
difficulties in obtaining good quality crystals and the
requirement of high pressure, the detailed study of the nature of
superconductivity remains a thrilling challenge.

Low pressure $\alpha$-Fe has a body center cubic (bcc) structure
and undergoes a martensitic transition to hexagonal (hcp)
$\varepsilon$-Fe for pressures higher than 12 GPa.\cite{Bancroft,
Bassett, Wang} According to Ref 9 and 10, the $\varepsilon$-Fe
phase is non-magnetic.\cite{Taylor, Nasu} Besides, it has been
reported that under pressure Fe loses its ferromagnetic character
due to the widening of the d band (i.e. a reduction in the density
of states), and then transforms into the hcp $\varepsilon$-Fe
phase, emphasizing the driving role of magnetism.\cite{Ekman,
Mathon}  The superconducting state emerges in this hexagonal phase
above 13~GPa and reaches a maximum $T_{c}$ of 2.2 K around 20~GPa
before disappearing at 31~GPa.\cite{Shimizu, Jaccard, Alex, Pablo}

The origin of Cooper pairing, whether it is mediated by phonons or
by  magnetic fluctuations still needs to be unveiled. Although
there has been no direct proof yet, the possibility of
electron-phonon (\textit{el-ph}) coupling is highly unlikely. The
rapid disappearance of superconductivity (SC) at 31~GPa compared
to the slower change of elastic properties (i.e. the
\textit{el-ph} coupling), and the presence of magnetic
fluctuations do not support this conjecture.\cite{Mazin}
Theoretical studies by Jarlborg \textit{et al.} have also
questioned the \textit{el-ph} coupling mechanism.\cite{Jarlborg}
Density functional theory calculations have predicted the
existence of the ordered antiferromagnetic (incommensurate spin
density wave) state in a small pressure region.\cite{Thakor}
Recently, evidence for weak magnetism, presumably
antiferromagnetic fluctuations, at pressures greater than 20~GPa
has been provided by x-ray emission spectroscopy.\cite{Monza}

The low temperature resistivity of $\varepsilon$-Fe has an unusual
temperature dependence $\rho(T)\sim A T^{5/3}$ up to at least $10
T_{c}$, with a  large value of coefficient $A$, which exhibits a
similar pressure dependence as the one of the superconducting
$T_{c}$.\cite{Alex, Pablo} SC is highly sensitive to crystal
disorder and the upper critical field H$_{c2}$ ($\sim$ 0.7 T) is
enhanced compared to the low superconducting $T_{c}$
value.\cite{Alex} These observations point towards an
unconventional nature of SC, mediated by spin fluctuations,
possibly of ferromagnetic nature.

In this paper, we report high pressure x-ray diffraction and
electric transport measurements on good quality Fe single
crystals. In order to address the question of the role of pressure
conditions (hydrostaticity) on the $\alpha\rightarrow\varepsilon$
transition of Fe, which was reported to be very sensitive to the
pressure medium,\cite{Taylor} the present resistivity
investigation was performed in a different pressure medium
(pyrophyllite) and is compared to  previous studies. Furthermore,
pressure cycling (increasing and decreasing) has been implemented
to check the effect on the transport properties near the
superconducting and magnetic/martensitic transitions. A broad
hysteresis is observed on pressure cycling in the room temperature
resistivity $\rho_{RT}$ (in agreement with x-ray diffraction) as
well as in low temperature transport parameters. Amazingly the
superconducting $T_{c}$ does not show a similar effect on pressure
cycling. This qualitative discrepancy is consistent with the
existence of a threshold residual resistivity for the occurrence
of the superconducting state, which is a hallmark of
unconventional SC. The transport parameters are analyzed in the
light of weakly ferromagnetic compound like ZrZn$_2$ \cite{Smith}
or the triplet superconductor Sr$_2$RuO$_4$ \cite{Mackenzie}.

\section{Experiments}
The single crystal diffraction study at high pressure was
performed on I15, the Extreme Condition Beamline at Diamond Light
Source, UK. A monochromatic beam ($E = 33.94$~keV) was focussed
onto a thin (24 $\mu$m) single crystal (whisker) placed in a
diamond anvil cell (DAC). The faces of the whisker were the
$\lbrace$100$\rbrace_{bcc}$ and the largest sample surface
(50$\times$34 $\mu$m$^{2}$) was perpendicular to the incident wave
vector. An area detector\cite{ATLAS} inclined by $5^{\circ}$ with
respect to the incoming  wave vector was used to collect the
single crystal images (exposure time of 1 second) while scanning
the $\phi$ axis. Daphne oil 7373 was used as a pressure medium and
the pressure was measured by the ruby fluorescence technique.

Resistivity measurements were performed on a Fe whisker with a
residual resistivity ratio RRR$\sim 250$. Our previous transport
measurements were initially made using steatite \cite{Alex, Pablo}
and subsequently Daphne oil \cite{Koshik} as pressure transmitting
media. From the width of the superconducting transition of the Pb
manometer the pressure gradients ($\Delta p/p$) in both media were
estimated to be about 5$\%$ and 3$\%$, respectively. In the
literature, the width ($w$) of the
$\alpha\leftrightarrow\varepsilon$ transition of Fe was reported
to be very sensitive to the pressure medium, ranging from $w\sim
0$ in helium to more than 10~GPa in a medium of very poor
hydrostaticity, like aluminium oxide. In spite of many efforts, we
could not succeed to increase sufficiently the maximum pressure of
our helium DAC for resistivity measurements.\cite{Holmes, Thesis}
Therefore we decided to try the opposite way and deliberately
chose to measure in pyrophyllite, a pressure medium with a
relatively low hydrostaticity. This modification was found to be
quite compatible with our standard technique where samples and the
Pb manometer are inserted in between two soft solid
disks.\cite{Jaccard2} Furthermore with the replacement of steatite
by pyrophyllite the pressure cell remained stable while releasing
the load, allowing us to cycle the pressure. In pyrophyllite, we
obtained $\Delta p/p\sim 0.08$. Pressure was changed at room
temperature and the resistivity of Fe was normalized to $\rho =
10.0\,\mu\Omega$cm at ambient conditions.\cite{Serway} Given that
the sintered diamond anvils of the Bridgman pressure cell are
slightly magnetic, special care was taken to obtain the correct
superconducting transition temperature of Pb and thus the
corresponding pressure inside the cell. An external low field coil
was used to compensate any remanent magnetic field of the anvil
cell.

\section{Results}
\subsection{X-ray diffraction}

Figure 1a and 1b are single crystal diffraction patterns of iron
just below and almost above its martensitic
$\alpha\rightarrow\varepsilon$ transition around 14~GPa. Each
pattern is the sum of 40 raw images corresponding to a $\phi$ scan
of 20$^{\circ}$ in the steps of $0.5^{\circ}$. With increasing
pressure there is a clear change in diffraction pattern and the
single crystal spots tend to become powder arcs. After two
pressure cycles (5 - 20~GPa) the patterns are almost completely
dominated by powder rings (not shown). The images shown in figure
1 correspond to the first pressurization.

\begin{figure}[btp]
\begin{center}
\includegraphics[scale=0.4]{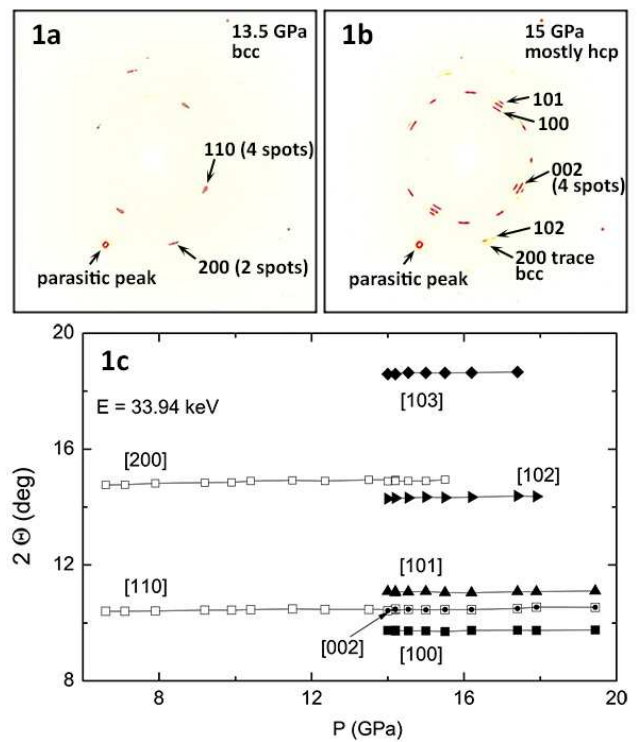}
\caption{(Color online)a) and b): Single crystal diffraction
patterns of iron just below and almost above its martensitic
$\alpha$ $\rightarrow$ $\varepsilon$ transition around 14~GPa. c):
Bragg angles of the diffraction spots or powder rings vs.
pressure.}
\end{center}
\end{figure}

As expected in the \textit{bcc} phase (Fig.~1a) there are four 110
and two 200 diffraction spots. The most interesting point in
Fig.~1b is that each 110$_{bcc}$ reflection changes into a
002$_{hcp}$ reflection. In addition, each 002$_{hcp}$ reflection
is followed by one 100$_{hcp}$ and one 101$_{hcp}$ reflection, and
additionally eight 101$_{hcp}$ reflections appear (Fig.~1b). Thus
the \textit{bcc} whisker transforms into four \textit{hcp} domains
related by the four-fold rotation along [100]$_{bcc}$ axis. To our
knowledge this is the first direct evidence of this well admitted
microscopic path of the martensitic transformation.\cite{Wang}
Fig.~1c shows the Bragg angles of the diffraction spots or powder
rings vs. $p$. It shows that the
\textit{bcc}$\rightarrow$\textit{hcp} transition starts a bit
below 14~GPa. From the spot intensities, it appears that
qualitatively a large fraction of the iron sample transforms in a
narrow pressure interval ($<$ 1~GPa) in agreement with the first
order character of the structural transition.\cite{Jephcoat}
However, there are weak traces of the 200$_{bcc}$ reflection up to
15.5~GPa and this allows us to roughly estimate the total
transition width $w\sim 1.5 - 2.0\,$GPa in agreement with the
literature.\cite{Mathon, Monza} The pressure dependence of the
110$_{bcc}$ reflection shows a smooth variation with $p$ and
becomes the 002$_{hcp}$ reflection. The 102 and 103 reflections of
the \textit{hcp} phase are very weak and undetectable beyond 17.4
or 17.9~GPa, respectively. For decreasing pressure the hcp phase
is observed down to pressures much lower than 14~GPa, and the
\textit{hcp$\rightarrow$bcc} transition occurs around 7~GPa with a
similar width as for increasing \textit{p}. Accordingly our
results confirm the large pressure hysteresis of 7~GPa observed in
previous studies.\cite{Taylor} For the second pressure cycle we
obtained the same values for the transition pressure and width.

\begin{figure}[tbp]
\includegraphics[scale=0.42]{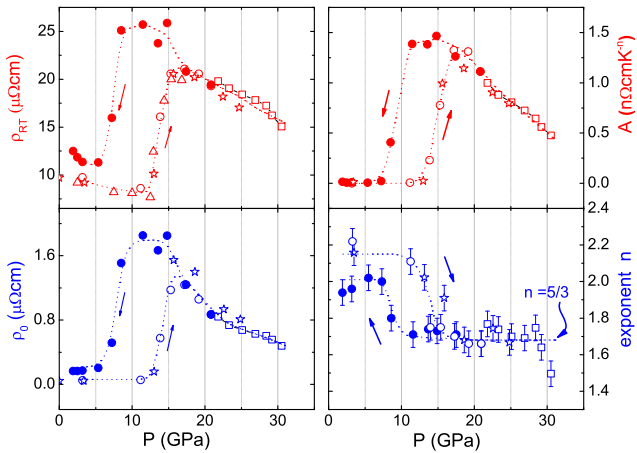}
\caption{(Color online) The pressure dependencies of $\rho_{RT}$
($T=290\,$K) and of the low temperature parameters $\rho_{0}$,
$A$, and $n$ show broad hysteresis while increasing (open circles)
and decreasing (filled circles) pressure. Open squares and  star
symbols correspond to the measurements performed in steatite
\cite{Alex, Pablo} and triangles to those performed in Daphne
oil.\cite{Koshik}}
\end{figure}

\subsection{Resistivity}

Following our previous studies on Fe in Daphne oil and steatite
media,\cite{Alex, Pablo} we performed electrical resistivity
measurements from room temperature down to 50\,mK and up to 21~GPa
using pyrophyllite as pressure medium. The normal state as well as
the superconducting properties in pyrophyllite were found to be
almost identical to those measured in other media. The resistivity
of $\alpha$-Fe is weakly pressure dependent. As a function of
temperature, $\rho(T)$ exhibits the typical properties of a
long-range ferromagnetic metal with large Curie temperature and
then varies superlinearly due to the addition of the
\textit{el-ph} and electron-magnon scattering terms. In
comparison, $\rho(T)$ of $\varepsilon$-Fe is strongly enhanced and
more pressure dependent. The residual resistivity $\rho_{0}$ is
increased by one order of magnitude and $\rho(T) = \rho_{0} +
AT^{n}$ with $n\simeq 5/3$ up to about 30\,K, and an enhanced
value of $A$. At higher temperatures $\rho(T)$ evolves towards a
nearly linear temperature dependence. We do not show these
$\rho(T)$ data here in order to avoid repetition. However we have
combined the results from previous measurements with the new data
to bring forth a consistent picture of the transport properties of
Fe.

Figure 2 shows the pressure variation of the room temperature
resistivity $\rho_{RT}$, as well as the low temperature parameters
$\rho_{0}$, $A$ and $n$ up to 30.5~GPa. Upon increasing $p$, our
recent measurements ($0\leq p \leq 21\,$GPa) match quite well with
the data obtained in steatite ($21\leq p \leq 30.5\,$GPa)
\cite{Pablo} as well as with previous data \cite{Jaccard, Alex,
Koshik}. The important point is that the pressure cell remained
quite stable when using the pyrophyllite medium and thus enabled
us to cycle the pressure. There are two main new results. First,
the resistivity as parameterized by $\rho_{RT}$, $\rho_{0}$, $A$
and $n$ shows a broad hysteresis of roughly $6.5\,$GPa around the
martensitic transition, in agreement with the x-ray diffraction
data. Second, with decreasing $p$, the hysteresis starts at about
19\,GPa which is the pressure of the maximum of the
superconducting transition temperature $T_{c}$.

Concerning $\rho_{RT}(p)$, enhanced magnetic scattering when
transiting from ferromagnetic $\alpha$-Fe to non-magnetic
$\varepsilon$-Fe leads to the increase in resistivity. The width
of the transition $w\sim 3\,$GPa, as observed in steatite is
slightly broader in pyrophyllite and narrower in Daphne oil. With
decreasing pressure, the $\varepsilon$-Fe phase persists with a
continuous rise in resistivity down to roughly $10\,$GPa, before
collapsing to $\alpha$-Fe. Similar to $\rho_{RT}$, $\rho_0$ also
shows a broad hysteresis and recovers low values for $p<3\,$GPa.
The $\rho_{RT}$ can be influenced by the change in the
\textit{el-ph} coupling, thus the hysteresis seen in $\rho_0$ is a
better signature of an intrinsic hysteresis at the magnetic
(martensitic) transition. This result is the first indication of a
hysteresis in the low temperature properties of iron.

The $A-$coefficient follows a similar trend as that of
$\rho_{RT}(p)$ and $\rho_{0}(p)$, showing a large increase at the
transition and then slowly decreasing in the $\varepsilon$-Fe
phase. The increase in $A(p)$ can be associated with the enhanced
spin fluctuations upon the transition to the $\varepsilon$-Fe
phase. Its large value evidences a strongly correlated phase and
supposedly the maximum observed at $19\,$GPa signals the location
of a quantum critical point (QCP). The extended $\varepsilon$-Fe
phase upon decreasing pressure leads to the increase in the $A$
value down to $\sim$ 12~GPa. The exponent $n$ also shows a
hysteresis with pressure cycling, going from $n$ $\sim$ 2.1,
characteristic of a long-range ferromagnet like $\alpha$-Fe, to
the more exotic value $n\approx 1.67\simeq 5/3$ in the
$\varepsilon$-Fe phase. The $n=5/3$ exponent indicates the
ferromagnetic nature of the spin fluctuations.\cite{Lonzarich} The
variation in $n(p)$ near the low pressure regime could be related
to the ferromagnetic domain wall scattering.

\begin{figure}
\includegraphics[scale=0.35]{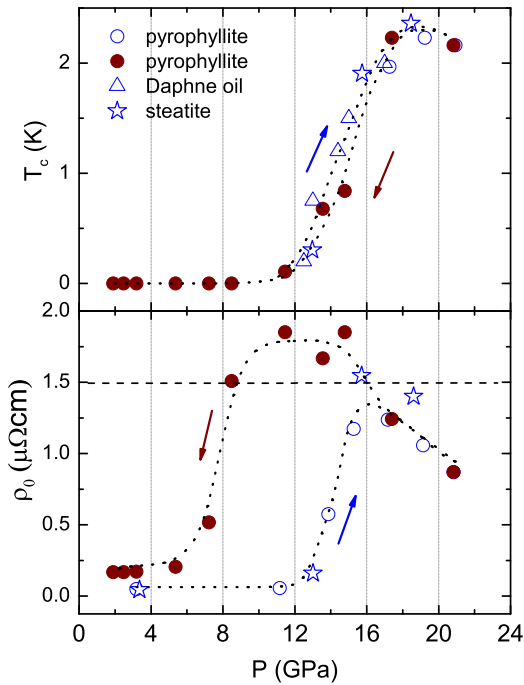}
\caption{(Color online) The pressure dependence of the
superconducting transition temperature $T_{c}^{onset}$ for
increasing (open symbols) and decreasing pressure (filled circles)
does not exhibit hysteresis. In combination with the $\rho_{0}(p)$
data (lower panel), this can be related to a strong suppression of
superconductivity in the $\varepsilon$-Fe phase beyond a certain
threshold $\rho_{0}$ value.}
\end{figure}

The top panel of Fig.~3 exhibits the pressure dependence of the
onset of the superconducting transition $T_{c}^{onset}$, where
$\rho(T)$ drops by $1\%$ of its lowest normal state value just
before transiting. With increasing pressure $T_{c}^{onset}$ is
first detected at $13\,$GPa, reaching a maximum value of $2.3\,$K
at $19\,$GPa, in good agreement with previous
reports.\cite{Shimizu, Jaccard, Alex, Pablo} However, $T_{c}(p)$
does not show a large hysteresis while decreasing pressure and it
is even lower around $15\,$GPa in comparison to the increasing
pressure data. Although the $\varepsilon$-Fe phase exists
prominently down to $\sim 10\,$GPa with a notably large
$A-$coefficient, $T_{c}(p)$ decreases sharply and vanishes at the
same pressure at which it had initially appeared. This behavior is
unexpected and at first sight it seems to contradict the view that
SC evolves concomitantly with the $A$ coefficient, suspected to
reflect the strength of the superconducting coupling in a spin
fluctuation scenario.\cite{Alex, Pablo} Nevertheless, such an
argument neglects the pair breaking effect due to the increase of
$\rho_{0}$ beyond $1.5\mu\Omega$cm while decreasing pressure, as
shown in the lower panel of Fig.~3.\cite{Alex, Thesis} The absence
of a hysteresis in $T_{c}(p)$ due to the increase of $\rho_{0}$
beyond $1.5\,\mu\Omega$cm is consistent with the notion of
unconventional SC in $\varepsilon$-Fe.

The temperature dependent part of the resistivity is plotted in
Fig.~4 against $T^{5/3}$ for increasing pressures between 15.3 and
$29.2\,$GPa. Excellent fits (dashed lines) are obtained up to a
temperature $T^{*}$, where data start to deviate upwards due to
the rapid rise of the \textit{el-ph} resistivity term. The slopes
of the fits are the $A-$coefficients shown in Fig.~2. In fact, the
$T^{5/3}$ law is accurately followed already from temperatures
just above $T_{c}$ (see different plots in Ref.~\onlinecite{Alex}
and \onlinecite{Pablo}) and then extends over more than an order
of magnitude up to $T^{*}$. It is also noteworthy that the
$T^{5/3}-$law is observed for pressures that cover almost the
entire superconducting domain, $13<p<31\,$GPa. Moreover, as shown
in the inset of Fig.~4, $T^{*}$ finds its maximum around 21~GPa,
i.e. close to the maxima of $T_{c}$, $A$ and $\rho_{0}$. Usually,
one expects $T^{*}$ $\propto$ \textit{A$^{-1/2}$} for a normal
Fermi liquid (n = 2), while in this case the higher $A$, the
higher $T^{*}$. Such a correlation, also observed in heavy
fermions or Fabre salts can be considered as an indication of a
QCP in $\varepsilon$-Fe in the vicinity of
20~GPa.\cite{Jaccard3,Sabina} In addition it seems unlikely that
the $T^{*}(p)$ maximum might be due to an artefact of the
$\textit{el-ph}$ term given that its pressure dependence is
expected to be monotone (see the discussion section).

\begin{figure}
\includegraphics[scale=0.42]{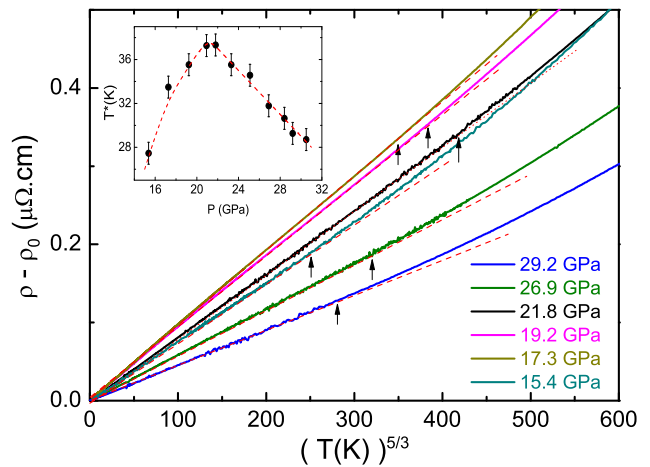}
\caption{(Color online) Temperature dependent part of resistivity
($\rho - \rho_{0}$) plotted as a function of $T^{5/3}$ for
selected pressures between 15.3 and 29.2~GPa. Dashed lines
correspond to the $AT^{5/3}$ fit. The inset shows the pressure
dependence of the temperature $T^*$ up to which the
$T^{5/3}$-dependence is observed.}
\end{figure}

Figure 5 shows the pressure dependence of the superconducting
$T_{c}$, estimated from three different criteria corresponding to
the  resistivity drop of 1$\%$, 10$\%$, and 100$\%$. To draw a
comprehensive $T_{c}-p$ phase diagram for Fe, the recent data
obtained in pyrophyllite are completed by previous measurements
done in steatite\cite{Alex, Pablo} and Daphne oil\cite{Koshik}.
Using the $1\%$ drop criterion ($T_{c}^{1\%}$), our results
confirm the bell-shape of $T_{c}(p)$, originally discovered by
Shimizu \textit{et al.}\cite{Shimizu} The pressure domain and the
maximum $T_{c}\approx 2.3\,$K are similar. For good samples (RRR
$\sim 200$ at $p = 0$) of different origins, all our results agree
without exception. Moreover, the $T_{c}^{1\%}$ values observed in
Daphne oil, steatite and pyrophyllite are in good agreement with
each other. A slight difference seems that the $\rho(T)$ drop is
somewhat more rapid in the best medium which is Daphne
oil\cite{Koshik}. The superconducting transition is very broad in
temperature and most often partial for all these media.
Considering a more restrictive criterion like $T_{c}^{10\%}$, the
superconducting region shrinks in $T$ and $p$, whereas the
complete ($>99\%$) $\rho(T)$ transitions are limited to a narrow
pressure domain between 19 and $23\,$GPa with maximum
$T_{c}^{100\%}$ of only $0.5\,$K. In fact, the $T_{c}(p)-$curve
exhibits a small asymmetry and its maximum in $p$ depends slightly
on the resistivity criteria (dashed line in Fig.~5). Both
$T_{c}^{100\%}(p)$ and $T^*(p)$ have maxima around $21\,$GPa. The
detection of complete resistive transitions strongly depends on
the measuring current or on the applied magnetic field, suggesting
the existence of superconducting islands with weak links. SC
starts to be suppressed for current densities $j$ as low as
$1\,$A/cm$^{2}$ or in magnetic fields of a few Gauss. Conversely
with the $T_{c}^{1\%}$ criterion, SC is much more robust. No
decrease of $T_{c}^{onset}$ was detected for $j =
10^{3}\,$A/cm$^{2}$ and a relatively high upper critical field
$H_{c2}(T\rightarrow 0)\approx 0.7\,$T was observed for such a low
$T_{c}$ metal. Let us add that small Meissner signals have been
reported\cite{Shimizu}, but we did not find any bulk signature of
SC by ac-calorimetry. The independence of results from the
pressure conditions strongly suggests that the $T_{c}^{1\%}(p)$
curve and in particular its rise above 12~GPa is intrinsic in
nature. Presumably, similar results would be obtained in solid
helium (i.e. in the pressure medium with the highest
hydrostaticity) because the very broad superconducting transition
comes mainly from the sample limitation and is not an experimental
artefact.

\begin{figure}[tbp]
\includegraphics[scale=0.4]{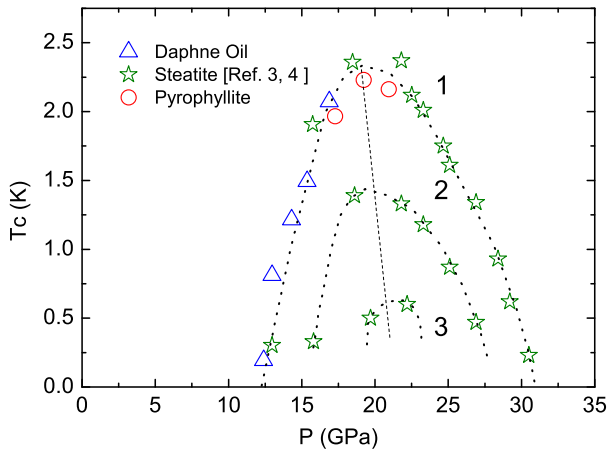}
\caption{(Color online) Superconducting $T_{c}$ versus pressure
phase diagram for Fe, measured in different pressure media. Dotted
curves marked as 1, 2, and 3 correspond to the transition
temperature $T_{c}$ taken from the 1$\%$, 10$\%$ and 100$\%$ drop
of resistivity from its normal state value, respectively.}
\end{figure}

\section{Discussion}

The x-ray diffraction measurements performed at room temperature
in Daphne oil pressure medium give a width $w\sim 1.5-2\,$GPa for
the $\alpha\leftrightarrow\varepsilon$ transition of Fe. The order
of the structural transition is not yet established since it is a
displacive transformation.\cite{Dupe} In comparison, the transport
measurements which probably reflect principally the magnetic
collapse indicate a larger width. For any $T\leq 300\,$K, the
resistivity (see Fig.~2) dramatically increases in the pressure
interval 12.5 - 15.5~GPa. Most likely only a small part of this
increase is due to the change of the \textit{el-ph}
coupling\cite{Alex}, as inferred from investigation of metastable
non-magnetic $\gamma$-Fe.\cite{Bohnenkamp} The width of the
transition $w\approx 3\,$GPa agrees with the value $w\sim
2.4\,$GPa observed by x-ray magnetic circular
dichroism.\cite{Mathon} Moreover, we find that $w$ is nearly the
same in Daphne oil, steatite or pyrophyllite media, i.e. weakly
dependent on the pressure conditions, in disagreement with Taylor
\textit{et al.}\cite{Taylor}  who reported different $w$ for
different pressure media. Our observations are consistent with a
width $w$ considerably larger than the respective $\Delta p$
inside the pressure cell (in the range $3\%<\Delta p/p<8\%$), and
indicate that $w$ is intrinsic to the
$\alpha\leftrightarrow\varepsilon$ structural and magnetic
transition. Thus the growth of anomalous scattering up to a
hypothetical QCP located around 19 GPa is a genuine property of
$\varepsilon$-Fe.

Interestingly, the room temperature resistivity $\rho_{RT}(p)$ has
a cusp at $13\,$GPa in steatite as shown by studies with small
pressure increments\cite{Jaccard}, and an even bigger cusp ($30\%$
jump) in Daphne oil. This sharp anomaly marks the start of the
breakdown of the long-range ferromagnetic order which slightly
precedes the structural transition by about
$0.5\,$GPa.\cite{Mathon} Moreover, as the emergence of SC
coincides with the cusp in $\rho_{RT}(p)$, the coexistence of SC
with ferromagnetic clusters seems clear at least up to $15\,$GPa.
At that pressure the exponent $n$ of the temperature power law of
resistivity is already locked to $n$ = 5/3, reflecting the
presence of ferromagnetic spin fluctuations. Aside from that it is
instructive to compare the behavior of Fe with Pb (our manometer),
which undergoes a martensitic $fcc \rightarrow hcp$ transformation
between 13 and $16\,$GPa.\cite{Kuznetsov} In this pressure window,
$\rho_{RT}$ increases smoothly  by around $20\%$ without any cusp.
At low temperature the superconducting resistive transition at
$T_{c}$ remains narrow and $T_{c}(p)$ does not deviate from its
slow decrease with increasing $p$. Apparently the phonon modes
responsible for the conventional SC in Pb are not affected by the
structural transition.

The most interesting result of the pressure cycling is that the
increasing and decreasing $p$ data merge only at $p_{max}\approx
19\,$GPa, suggesting that the $\alpha\rightarrow\varepsilon$
magnetic transition has a tail and that non-magnetic
$\varepsilon$-Fe is realized only for $p\geq p_{max}$. This is
true for the four quantities shown in Fig.~2, but not for $T_{c}$
presumably due to a sharp pair breaking effect. For instance
considering the $A-$coefficient, above 12.5~GPa where the
transition starts, the difference between the decreasing and
increasing $A(p)$ values can be viewed as directly linked to the
amount $\eta$ of magnetic clusters, remnant of the ferromagnetic
$\alpha$-Fe. The scenario is that these magnetically unstable
clusters induce ferromagnetic fluctuations which grow up to a QCP
marked by the vanishing of $\eta$ at \textit{$p_{max}$}. As a
result at the QCP the resistivity is maximum and in particular the
coefficient $A$ as well as the superconducting $T_{c}$.
Furthermore the $n$ = 5/3 temperature power law of resistivity
extends up to a maximum $T^{*}$ at almost the same $p$. It is
noteworthy that, at a pressure close to \textit{$p_{max}$}, a cusp
has been reported in the weak magnetic signal detected by x-ray
emission spectroscopy.\cite{Monza} However, such a feature could
also be related to other types of electronic instabilities like an
electronic topological transition.\cite{Glazyrin} With decreasing
$p$, the strength of the interaction between the electrons and
spin fluctuations is maximum at about 13~GPa where $A$ takes its
maximum, indicating that the electronic instability has the same
hysteresis as the structural transition. This electronic
instability appears to be a precursor sign of the long range
ferromagnetic order which becomes stable around 7~GPa below the
instability. The decrease of $A$ at lower $p$ would be due to the
progressive growth of ferromagnetically stable clusters on
approaching the bcc phase. Up to now it is not clear why the total
width of the magnetic transition including its tail corresponds to
the observed broad hysteresis of 7~GPa, but our observation
supports the driving role of magnetism in the
$\alpha\leftrightarrow\varepsilon$ transition of Fe. With
increasing $p$, the value of the $A$ coefficient appears to track
$T_{c}$, implying that the same ferromagnetic fluctuations
responsible for the non-Fermi liquid behavior in resistivity may
also be responsible for the superconducting pairing interaction.
Moreover, reaching 31~GPa, the $A$ coefficient seems to fall below
a certain minimum threshold value, necessary for SC. However, this
point is less clear for the emergence of $T_{c}$ around 13~GPa,
simply because the $A(p)$ and $T_{c}(p)$ variations are too rapid
and likely to be smeared by the $p$ gradient.

The absence of hysteresis in $T_{c}(p)$ (Fig.~3) suggests the
existence of a certain $\rho_{0}-$value, beyond which SC is
suppressed. Indeed a strong enhancement of $\rho_{0}$ is observed
in the hcp phase, mimicking the one seen in $A(p)$. As to its
origin, pressure cycling may induce some micro-structural changes
leading to a slow decline of the single crystallinity, as can be
inferred from the x-ray diffraction data. However, these changes
are not very significant, at least in affecting $\rho_{0}$, since
it finally recovers to low values at low pressure. As an
alternative explanation, we suggest that the effect of lattice
disorder on $\rho_{0}$ gets substantially amplified by spin
fluctuations in this particular pressure region, hence leading to
the observed enhancement in $\rho_{0}$. Coming back to an eventual
threshold value of $\rho_{0}$ for SC, such a phenomenon is also
found for example in in the pressure-induced superconductor
CePd$_{2}$Si$_{2}$.\cite{Raymond} Actually, the best documented
case is the spin triplet superconductor Sr$_2$RuO$_4$ for which
non-magnetic impurities kill the superconducting state when the
carrier mean free path $l\propto\rho_{0}^{-1}$ falls below the
superconducting coherence-length $\xi$. Mackenzie \textit{et al.}
have shown that the generalized theoretical model for non-magnetic
impurities in an unconventional superconductor (which is based on
the pair breaking Abrikosov-Gorkov theory for magnetic impurities
in BCS superconductors), fits very well with the dependence of
$T_{c}(\rho_{0})$.\cite{Mackenzie} A threshold of $\rho_{0}$ = 1.1
$\mu \Omega$cm was established for Sr$_2$RuO$_4$ samples of
different chemical purities. For Fe, when the impurity level is
below $100\,$ppm the crucial parameter is not the chemical purity
but the metallurgical state of the sample.\cite{Alex} The
threshold $\rho_{0}=1.5 \mu\Omega$cm was estimated by controlling
the intrinsic sample disorder, either by rolling (cold work
induces dislocation defects) or by annealing. The electronic mean
free path \textit{l} $\propto$ $\rho_{0}^{-1}$ has a threshold
value around 10 nm for SC. According to the critical field data,
the coherence length $\xi$ appears to be close to \textit{l}, i.e.
the clean limit is required which supports an unconventional
nature for the paring mechanism. For Sr$_2$RuO$_4$ a narrow
transition is observed at $T_{c}$ when $\rho_{0}$ is much lower
than the threshold value. This condition is never satisfied in Fe
and thus only broad transitions are observed. Moreover, when
$T_{c}$ decreases, the criterion $\xi<l$ introduces further
limitations because $T_{c}\propto \xi^{-1}$. Obtaining narrow
resistive transitions would be essential in order to progress in
the study of SC of Fe. However, there is little hope for that as
the $\rho_{0}$ enhancement when entering the $\varepsilon$-Fe is
in a large part intrinsic, i.e. only a small decrease is observed
with improving sample quality. Also, the \textit{in-situ}
annealing of the sample seems impossible. Iron samples with a
sufficiently low $\rho_{0}$ should exhibit bulk SC in the pressure
domain $13<p<31\,$GPa with a maximum $T_{c}-$value higher than
$2.5\,$K.

The power law $\rho(T)\propto A T^{5/3}$ has been reported for
some weakly ferromagnetic metals including ZrZn$_{2}$,\cite{Smith}
Ni$_{3}$Al,\cite{Niklowitz} and Pd$_{x}$Ni$_{1-x}$.\cite{Nicklas}
In the case of the alloy Pd$_{x}$Ni$_{1-x}$, a ferromagnetic
quantum critical point clearly occurs for $x = 0.025$ where $n =
5/3$ is minimum while $A = 2 {\rm n}\Omega\,{\rm cm}/{\rm
K}^{5/3}$ is maximum, culminating at a value a bit larger than
that of Fe at $p_{max}\approx 19\,$GPa. For ZrZn$_{2}$ the picture
is less standard: surprisingly $A = 9 {\rm n}\Omega\,{\rm cm}/{\rm
K}^{5/3}$ and $n\sim 5/3$ are almost $p-$independent up to
pressures close to $p_{c}=2\,$GPa, where the ferromagnetism is
suppressed completely and the exponent drops to $n\sim 3/2$.
Moreover, there is a change of slope at the Curie temperature in
the $T^{5/3}$ plot of the resistivity. These anomalies have been
considered to be compatible with the marginal Fermi liquid state
expected in weakly ferromagnetic metals. In the case of Fe the
situation is still different as $n$ is fixed on a broad $p-$range
outside the ferromagnetic phase while $A(p)$ varies strongly.

The subtraction of a phonon term $\rho_{ph}$ to the total
resistivity (data from Ref.~\onlinecite{Pablo}) suggests that the
$T^{5/3}$ temperature dependence might hold up to $T\sim 200\,$K,
i.e. a temperature much higher than $T^{*}$ as defined in Fig.~4.
However, extension of such an analysis to pressures below the
superconducting $T_{c}(p)$ maximum leads to an unlikely pressure
dependence of $\rho_{ph}$. Furthermore, the data treatment assumes
a strict validity of Matthiessen's rule considering that $A
T^{*5/3}$ is only about 30$\%$ of $\rho_{0}$ and that the pressure
in our cell is sufficiently temperature independent, which seems
not to be the case. Indeed, the deviation from linearity of the
resistivity $\rho(T)$ of Pb points to a slight increase of
pressure above 80 K (by about 5$\%$ up to 300 K) and $p$ can be
considered as constant only below 50 K. Therefore the simple
$T^{5/3}$ plot of Fig.~4 is the most reliable analysis, showing
the occurrence of the $T^{*}(p)$ maximum. Nevertheless, the
resistivity term ascribed to spin fluctuations persists up to 300
K with an unknown $T$ dependence that is not far from $T^{5/3}$.
It is also noteworthy that we did not observe any anomaly which
could mark a Curie temperature similar to ZrZn$_{2}$. Accordingly,
resistivity measurements above 300 K are desirable in order to
evaluate the spin fluctuation temperature $T_{SF}$ which sets the
overall scale for spin mediated SC. For Fe a huge $T_{SF}$ seems
not to be excluded, explaining qualitatively the relatively high
superconducting $T_{c}$ value.

\section{Conclusions}

X-ray diffraction and electric transport measurements have been
carried out under high pressure on high quality Fe single
crystals. The x-ray data yield the first direct experimental
evidence of the microscopic path of the martensitic
$\alpha\leftrightarrow\varepsilon$ transformation. Combining this
study with previous ones, only a very weak dependence on the
pressure conditions is revealed. As a main outcome, it is now
evident that the superconducting pocket observed at the border of
ferromagnetic bcc-Fe is intrinsic to the hcp-Fe phase. As to its
origin, new insight comes from the unprecedented pressure cycling
of electric transport, and its analysis in terms of $\rho(T) =
\rho_{0} + AT^{n}$. Indeed, maxima in $A(p)$ and $\rho_0(p)$ are
observed (as well as $n\approx 5/3$) slightly above the structural
transformation (i.e. within the hcp phase), with a similar
hysteresis in pressure. These features likely signal a region of
strong ferromagnetic fluctuations, which may as well be
responsible for superconductivity, since $T_c(p)$ culminates in
the same pressure range. As a synoptic scenario, we suggest that
the magnetic transition has a tail (of a yet unknown nature)
ending at a QCP or another type of electronic instability,
precisely where the ferromagnetic spin fluctuations are maximum.
Given the proximity to long-range ferromagnetic order, it may act
as its precursor sign. The striking absence of hysteresis in
$T_c(p)$ may be explained by the high sensitivity of $T_c$ on
$\rho_{0}$ and the electronic mean free path, which additionally
points to an unconventional nature of the superconducting state.
Further experimental and theoretical progress is still necessary
to understand in detail the microscopic interplay between the
$\alpha\leftrightarrow\varepsilon$ structural and magnetic
transitions in elementary Fe, in particular in order to unveil the
nature of the electronic instability inside the hcp phase.
Concerning superconductivity, experimental improvements (such as
narrow resistive transitions) seem however compromised by the
intrinsic rise of $\rho_{0}$ and still represent an enormous
challenge.

Acknowledgements: We thank I. Ynada, H. Kohara and Y. Onuki for
providing Fe whiskers, J. Flouquet for useful discussions, M.
Lopes for technical assistance, and the Swiss National Science
Foundation for financial support.

\end{document}